\documentclass[sigplan,noacm]{acmart}
\usepackage{svg}
\usepackage{balance}
\usepackage{booktabs}
\usepackage{subcaption}
\usepackage{multirow}
\usepackage{listings}
\usepackage{xcolor}
\usepackage{tikz}
\usepackage[ruled,vlined,linesnumbered]{algorithm2e}
\PassOptionsToPackage{hyphens}{url}\usepackage{hyperref}

\AtBeginDocument{%
  }

\begin{document}

\title{PoisonCap: Efficient Hierarchical Temporal Safety for CHERI}

\author{Yuecheng Wang}
\orcid{0009-0009-8075-7006}
\affiliation{%
  \institution{University of Cambridge}
  \country{Cambridge, United Kingdom}
}
\email{yuecheng.wang@cl.cam.ac.uk}
\authornote{Corresponding author}

\author{Jonathan Woodruff}
\orcid{0000-0003-3971-2681}
\affiliation{%
  \institution{University of Cambridge}
  \country{Cambridge, United Kingdom}
}
\affiliation{%
  \institution{Capabilities Limited}
  \country{Cambridge, United Kingdom}
}
\email{jonathan.woodruff@cl.cam.ac.uk}

\author{Alfredo Mazzinghi}
\orcid{0009-0009-2941-4729}
\affiliation{%
  \institution{University of Cambridge}
  \country{Cambridge, United Kingdom}
}
\affiliation{%
  \institution{Capabilities Limited}
  \country{Cambridge, United Kingdom}
}
\email{alfredo.mazzinghi@cl.cam.ac.uk}

\author{Peter Rugg}
\orcid{0009-0000-2976-0474}
\affiliation{%
  \institution{University of Cambridge}
  \country{Cambridge, United Kingdom}
}
\affiliation{%
  \institution{Capabilities Limited}
  \country{Cambridge, United Kingdom}
}
\email{peter.rugg@cl.cam.ac.uk}

\author{Alexandre Joannou}
\orcid{0000-0002-3161-2638}
\affiliation{%
  \institution{University of Cambridge}
  \country{Cambridge, United Kingdom}
}
\affiliation{%
  \institution{Capabilities Limited}
  \country{Cambridge, United Kingdom}
}
\email{alexandre.joannou@cl.cam.ac.uk}

\author{Samuel W. Stark}
\orcid{0000-0002-7268-9471}
\affiliation{%
  \institution{University of Cambridge}
  \country{Cambridge, United Kingdom}
}
\email{samuel.stark@cl.cam.ac.uk}

\author{Robert N. M. Watson}
\orcid{0000-0001-8139-8783}
\affiliation{%
  \institution{University of Cambridge}
  \country{Cambridge, United Kingdom}
}
\affiliation{%
  \institution{Capabilities Limited}
  \country{Cambridge, United Kingdom}
}
\email{robert.watson@cl.cam.ac.uk}

\author{Simon W. Moore}
\orcid{0000-0002-2806-495X}
\affiliation{%
  \institution{University of Cambridge}
  \country{Cambridge, United Kingdom}
}
\affiliation{%
  \institution{Capabilities Limited}
  \country{Cambridge, United Kingdom}
}
\email{simon.moore@cl.cam.ac.uk}

\renewcommand{\shortauthors}{Yuecheng Wang et al.}

\begin{abstract}
In this paper, we present PoisonCap: scalable temporal safety with strict use-after-free protection and initialization safety for CHERI systems.
Efficient memory safety is an increasing priority for programming languages, operating systems, and hardware designs, and CHERI is a leading hardware/software system that provides native spatial safety and a foundation for temporal memory safety~\cite{cheriv9}.
Cornucopia Reloaded, the current state-of-the-art CHERI temporal safety solution, provides \emph{use-after-reallocation} safety but does not enforce immediate access denial following free() and is not able to enforce initialization safety~\cite{Cornucopia_reloaded}.
We show that a new \emph{poison} capability format can be used to enforce strict use-after-free and initialization safety, and also to communicate memory state to the microarchitecture for efficient cache management of quarantined memory.
We enable elegant delegation of memory poisoning privilege using capability bounds to allow nested allocators to enforce safety on their consumers without disturbing upstream allocators.
PoisonCap can replace the Cornucopia shadow bitmap and also automatically zeros memory on reallocation or optionally traps on read-before-write to enforce initialization safety.
As a result, PoisonCap strengthens CHERI temporal safety without
additional performance overhead relative to a Cornucopia baseline that zeros memory before reallocation.
\end{abstract}




\maketitle
\pagestyle{plain}

\section{Introduction}
Memory safety enforcement is a cornerstone of software security. 
Operating systems, programming languages, and CPU implementations are now built with support for preserving memory safety to mitigate security vulnerabilities;
examples include OS page zeroing, Rust, Java, ARM PAC, and ARM MTE~\cite{ArmCPUSecurity,ARM_MTE}.
CHERI is a promising technology that extends addresses with hardware-protected bounds and permissions so that every memory access is within the bounds of the original allocation.
These unforgeable, bounded pointers are called \emph{capabilities}, as they grant a program the capability to access a memory location.
CHERI, therefore, enforces \emph{spatial safety} in hardware, but does not natively prevent memory accesses through pointers that have been passed to \texttt{free()}.
Nevertheless, CHERI primitives have been leveraged to enforce \emph{temporal safety} to some extent.
Cornucopia Reloaded~\cite{Cornucopia_reloaded}, the state-of-the-art CHERI temporal safety solution, quarantines freed allocations until a sweep of the address space revokes all dangling heap capabilities,
similar to other deferred-reclamation approaches~\cite{ainsworthMarkUsDropinUseafterfree2020, erdosMineSweeperCleanSweep2022}.
The primary safety shortcomings of Cornucopia Reloaded are the lack of strict use-after-free detection and the lack of initialization safety.
That is, programs may mistakenly access memory allocations after calling ~\texttt{free()} without being detected or may depend on the value of fresh allocation data that is not guaranteed to be initialized.
Strict use-after-free enforcement detects program bugs that could be masked by delayed revocation.
Initialization safety contributes substantially to CHERI security; 
a comprehensive analysis of CHERI v7 by the Microsoft Security Response Center (MSRC) in 2019 found that initialization safety was a large fraction of the patched vulnerabilities not covered by CHERI at the time.
While over 55\% of their patched vulnerabilities would have been mitigated by CHERI spatial and temporal safety, an additional 12\% were caused by uninitialized access~\cite{cheri_security_analysis}.
As indicated by MSRC in this report, zero initialization is not sufficient to enforce initialization safety as this still leads to undefined behavior~\cite{msrc_uninitialised}.

In addition to these security shortcomings, Cornucopia Reloaded does not naturally support nested allocators, and wastes cache on caching quarantined allocations.

We propose PoisonCap: a hardware-software co-design that enforces deterministic use-after-free mitigation and initialization safety on application-class CHERI systems while communicating liveness to the cache hierarchy to improve performance.
PoisonCap introduces an architectural extension to CHERI that we call \emph{poison capabilities}.
Poison capabilities are a new capability type that prevents the memory in which they are held from being accessed.
PoisonCap replaces memory zeroing with memory poisoning, enabling immediate prevention of use after free and read-before-write for fresh allocations while making the out-of-use status of memory visible to the cache subsystem.
While PoisonCap still relies on revocation, the architectural poison mechanism can replace Cornucopia's shadow bitmap, a software approach to marking freed memory, enabling dangling pointer detection during the sweep without memory overhead.

PoisonCap also enables hierarchical use-after-free and uninitialized access protection for nested allocators.
ARM MTE~\cite{ARM_MTE} and CHERIoT~\cite{CHERIoT} track allocation status in a single hardware-maintained shadow space, forcing system designers to choose exactly one layer to use the mechanism.
PoisonCap uses CHERI capability bounds metadata to enforce memory poisoning hierarchically so that more privileged layers can freely manipulate memory poisoned by lower layers.

We have added memory poisoning support in the CheriBSD, userpace software stack and kernel revoker, proving compatibility with a full POSIX operating system.
PoisonCap delivers the following contributions:
\begin{itemize}
    \item We introduce a poison capability format to enforce strict use-after-free protection and initialization safety on CHERI.
    \item We use bounds in poison capabilities to scale protection across multiple allocation layers while being compatible with the current software stack.
    \item We demonstrate a prototype revocation mechanism that uses PoisonCaps to identify dangling capabilities in place of Cornucopia's shadow bitmap, enabling hierarchical revocation and potentially reducing revocation metadata overhead.
    \item We implement PoisonCap extensions in the CHERI RISC-V QEMU full system emulator~\cite{qemu} and in the CHERI-Toooba CPU~\cite{ruggSuiteProcessorsExplore2024}, a superscalar, out-of-order FPGA soft-core.
    \item We extend the CHERI-Toooba memory system to manage unused memory flows in the microarchitecture to improve cache efficiency.
    \item We add support for poison capabilities in the CHERI-enabled Clang/LLVM compiler~\cite{llvm} and CheriBSD OS~\cite{cheribsd}. 
    \item We evaluate poison capabilities using NIST Juliet test suite and MSET~\cite{Juliet, vintila2025mset} for security, and using SPEC CPU2006 INT and PARSEC for performance~\cite{henningSPECCPU2006Benchmark2006, bienia2008parsec}. 
\end{itemize}

\section{Background}
PoisonCap addresses long-standing challenges in computer systems that are currently mitigated by an array of software and hardware mechanisms, including both commercial deployments and academic proposals.
These challenges lie at the intersection of security, composability, and memory management in both programs and processor microarchitecture.

\subsection{CHERI}

CHERI is an architectural extension in which C/C++ pointers are
represented as capabilities rather than unprotected integer addresses.
In addition to an address, a capability encodes bounds and permissions and carries a hardware-maintained validity tag. Before completing a memory access, the processor checks the dereferenced capability's tag, sealing state, permissions, and bounds. These checks are efficient, as all required metadata is encoded directly in the capability, with no external metadata.

CHERI follows a decentralized capability model: capabilities can be copied and distributed throughout a program without registering them in, or checking them against, a central authority table. 
Capability derivation is also \emph{monotonic}: software may restrict a capability's authority, for example by narrowing its  bounds, but cannot subsequently increase that authority. Thus, a parent allocation capability can be narrowed to represent a child allocation, while the derived capability cannot subsequently be widened to recover the broader authority of the parent capability.

\subsection{CHERI Temporal Safety Systems}
\label{subsec:Cornucopia}

CHERI temporal safety systems for C and C++ have sought to enforce
lifetime duration for explicitly allocated heap objects.
C and C++ programs explicitly manage allocated storage duration using
the \texttt{alloc()} and \texttt{free()} family of functions.
Furthermore, well-behaved programs will execute `destructors' before objects are \texttt{free()}-d, placing them into an invalid state.
As with Address Sanitizer~\cite{AddressSanitizer}, CHERI temporal safety solutions
use the following definitions for enforcing temporal safety:
\begin{itemize}
    \item \bf{Freed Allocation}: An allocation whose reference has been passed
    to the \texttt{free()} function by the program.
    \item \bf{Dangling Pointer}: A pointer in memory that references a freed allocation.
    \item \bf{Illegal Access}: An access through a dangling pointer to a freed allocation.
\end{itemize}
As a result, the system may regard any access to an object after the program's call to \texttt{free()}
as dangling and liable to fail even if enforcement is only guaranteed before reallocation.
Pointers to objects in quarantine are considered dangling;
indeed, Address Sanitizer quarantines freed memory and traps access to these quarantined allocations.

CHERI's decentralized design makes capability revocation challenging.
When software requests the deallocation of heap objects, for example by
calling \texttt{free()}, copies of capabilities referring to the
allocation may remain distributed throughout memory. Ordinary
CHERI access checks cannot identify capabilities referring
to these deallocated objects.

Several CHERI systems implement capability revocation to address this problem.
CHERIvoke~\cite{CHERIvoke} introduced batched sweeping revocation; 
storage awaiting reclamation is quarantined and recorded in a
revocation bitmap until a sweep invalidates capabilities referring to
it.
Cornucopia~\cite{wesleyfilardoCornucopiaTemporalSafety2020} and
Cornucopia Reloaded~\cite{Cornucopia_reloaded} develop this approach
into concurrent, kernel-supported revocation for general-purpose CHERI
systems. For embedded systems, CHERIoT~\cite{CHERIoT} combines quarantine,
bitmap-based revocation, and sweeping, but additionally uses a hardware
load filter to check capabilities against the revocation bitmap as they
are loaded.

\textbf{Cornucopia Reloaded.}
PoisonCap builds directly on Cornucopia Reloaded, so we will describe its reclamation model in more detail.
A call to \texttt{free()} does not immediately make the allocation's storage available for reuse.
Cornucopia instead places the storage in quarantine and records the allocation in a shadow bitmap. While quarantined, the storage is not available for reallocation. 
Once quarantined memory reaches a configurable threshold, a revocation sweep scans the address space and uses the shadow bitmap to identify and invalidate capabilities referring to quarantined allocations. Only after revocation is the storage reclaimed and made available to the allocator for reuse.

Cornucopia therefore preserves CHERI's decentralized capability model while using quarantine and an external shadow bitmap to support revocation before reuse. 
This provides the baseline for PoisonCap, which instead represents revocation state using tagged poison values stored directly in memory.

\subsection{Memory Zeroing}
Zero initialization is a common strategy to mitigate uninitialized memory access and information leakage vulnerabilities.
By zeroing freed memory, the data (and capabilities) belonging to a previous allocation are cleared, and memory is also initialized to a predictable value prior to being accessed.

Nevertheless, CHERI research prototype allocators have been found to lack memory zeroing, presumably due to performance overhead, allowing capability leakage to heap ``scavengers''~\cite{CapLeakage}.
We assume that a properly implemented CHERI allocator may zero quarantined memory to prevent disclosure of previous allocation contents. 
Under Cornucopia Reloaded, stale capabilities may remain usable during quarantine, so zeroing on \texttt{free()} prevents such capabilities from observing the previous contents of the allocation.
Zeroing on reallocation additionally provides zero initialized storage to the subsequent allocation.

While allocation zeroing is certainly best-practice for current CHERI systems, MSRC points out that even zeroing is not sufficient for initialization safety~\cite{msrc_uninitialised}, which may lead to undefined program behavior. 
Zeroing also does not communicate ``liveness'' of data to the caches.
Zeroing on free to prevent leakage will bring lines into the cache; this wastes cache capacity, as these lines are now ``dead''.
Whether memory is zeroed on free or not, caches are polluted with dead data, either zeroed or abandoned, and DRAM bandwidth is wasted for ``dead'' data that program logic must not depend on.

\subsection{Nested Allocators}
\label{subsec:nested_allocators}
\begin{table*}[t] 
\caption{Examples of different types of CVEs arise in memory allocators}\label{table:nested_cwes}
\begin{tabular}{lccccccc}
\toprule
&  &  &  &  & \multicolumn{2}{c}{Mitigated by} & \\ \cmidrule{6-7}
& Vulnerability & Allocator & Memory Root & Required Safety & CHERI & PoisonCap & Source \\
\midrule
Postgres&buffer overflow& AllocSet~\cite{ExploringPostgressArena}&malloc&spatial&Yes&Yes&cve-2026-2007~\cite{CVE-2026-2007}\\
Postgres&nested use-after-free& AllocSet~\cite{ExploringPostgressArena}&malloc&temporal&No&Yes&bug report~\cite{gBugHeapUse2024}\\

Nginx &buffer overflow& ngx pool~\cite{nginx}&malloc&spatial&Yes&Yes&cve-2022-41741~\cite{CVE-2022-41741}\\
Apache &nested use-after-free & apr pool~\cite{ApachePortableRuntime}&malloc&temporal&No&Yes&cve-2019-0196~\cite{cve-2019-0196}\\
OpenSSL &out-of-bounds read &freelists~\cite{RootCauseStruct}&malloc&spatial&Yes&Yes&HeartBleed~\cite{CVE-2014-0160} \\
ProFTPD&nested use-after-free&memory pool~\cite{ProFTPDpool}&malloc&temporal&No&Yes&cve-2020-9273~\cite{CVE-2020-9273}\\
OpenJPEG &uninitialized access& libc allocator&mmap&initialization&No&Yes&cve-2025-54874~\cite{CVE-2025-54874}\\
libjxl&uninitialized access&libc allocator&mmap&initialization&No&Yes&cve-2025-12474~\cite{CVE-2025-12474} \\
\bottomrule
\end{tabular}
\end{table*}

As the complexity of the modern software stack increases, programming languages and user-space applications have started to implement custom allocators to optimize performance.
The custom allocators are often \emph{nested} allocators~\cite{pymalloc, nginx, ApachePortableRuntime,ExploringPostgressArena, MEMSYS5}, sub-allocating memory provided by the libc allocator.

Caching allocations of a particular type is a common use-case for custom allocators;
OpenSSL and Apache maintain \emph{freelists}, memory pools of freed memory that have not been freed back to the libc memory allocator to facilitate swift reallocation~\cite{ApachePortableRuntime,RootCauseStruct}. The freelist strategy has exposed OpenSSL to security vulnerabilities~\cite{NVDCve20105298,CVE-2014-0160}, as listed in Table \ref{table:nested_cwes}, including the Heartbleed vulnerability~\cite{CVE-2014-0160}.
For this reason, the freelist is disabled by default by the mainstream OpenSSL and its Google-maintained fork BoringSSL~\cite{SSLShuoChens} to enable system allocator protections.

It has long been challenging to detect memory vulnerabilities across nested allocation layers, as many software defenses are designed primarily for system allocators and therefore require explicit integration with application-specific sub-allocators.
This is because applications' nested internal allocation and deallocation operations are hidden behind a large allocation obtained from \texttt{malloc}~\cite{allocatorhidesbug}.

Ideally, memory safety mechanisms would nest elegantly such that multiple layers of allocation could rely on the same safety mechanism.
This is true of CHERI spatial safety, as nested allocators can naturally subset the bounds of a capability, achieving safety at every layer.
Temporal safety would benefit from a similar nested mechanism, as dangling capabilities may arise in any allocation layer.
Ideally, temporal safety would be enforced hierarchically so that memory freed at each allocation level would become immediately unavailable at that level while not preventing access through references in a parent allocator.

Unfortunately, most hardware and software temporal safety mitigation only support enforcement for a single layer of allocation~\cite{ainsworthMarkUsDropinUseafterfree2020, erdosMineSweeperCleanSweep2022, CHERIoT, Cornucopia_reloaded, ARM_MTE, AddressSanitizer}.
Application-class and embedded CHERI systems maintains a shadow bitmap data structure to support revocation, which is typically used in the libc allocation layer.
One bit in this data structure corresponds to a freed word in the heap~\cite{Cornucopia_reloaded, CHERIoT}, incurring at least 1/128 overhead for the heap.
While this overhead might be acceptable if CHERI is only targeted to support use-after-free in the libc allocation layer, using this same mechanism in nested allocation layers would likely require multiple shadow bitmaps, increasing memory overhead.

\subsection{Delayed Reallocation}
State-of-the-art memory allocators in unmanaged languages, e.g.\ Jemalloc~\cite{Jemalloc} and Snmalloc~\cite{snmalloc}, immediately reallocate freed memory to improve performance, as recently freed memory is more likely to reside in the cache than fresh memory allocations.
This property exposes software to UAF (use-after-free) vulnerabilities when dangling pointers persist.
Recent temporal memory safety approaches include memory tagging~\cite{AddressSanitizer,ARM_MTE}, deferred-reclamation and garbage-collection-inspired schemes that delay memory reuse until stale references have been revoked or become unreachable~\cite{ainsworthMarkUsDropinUseafterfree2020, erdosMineSweeperCleanSweep2022, Cornucopia_reloaded}, and one-time allocation~\cite{OTA}.
These approaches make different trade-offs among detection semantics, performance, memory overhead, and software compatibility.

Previous CHERI temporal safety enforcement relies on delayed reallocation~\cite{wesleyfilardoCornucopiaTemporalSafety2020, Cornucopia_reloaded}.
Cornucopia Reloaded does not sweep memory immediately on free to revoke dangling capabilities, but places freed memory into a quarantine buffer, not yet returning it to the allocator~\cite{Cornucopia_reloaded}.
This batching strategy significantly reduces the overhead of memory sweeping while ensuring that reclaimed memory cannot be reused until stale capabilities have been revoked.
During quarantine, existing stale capabilities may remain usable because Cornucopia does not enforce access denial at the program-visible free() boundary; accesses through stale capabilities may continue until revocation.

In addition to delaying fault detection, delayed reallocation also introduces performance overheads.
Modern user-space allocators such as Jemalloc and Snmalloc are optimized for reusing recently freed memory to exploit cache locality~\cite{Jemalloc,snmalloc}.
Indeed, we have observed a higher cache miss rate and additional DRAM traffic while delayed reallocation is enabled.

\subsection{Embedded Tagging vs. External Tagging}

Some commercial systems, such as ARM MTE and Apple EMTE~\cite{ARM_MTE,nohARMMTEPerformance2026}, use memory tags to identify freed objects.
ARM MTE requires storing and managing tags that are stored externally to memory data, and this often introduces additional physical memory overhead and disturbs physical memory accesses with table lookup. Moreover, memory tagging is often limited in tag size, which limits itself from being able to support temporal memory safety in a deterministically secure manner.  

PoisonCap does not require any external tagging beyond the CHERI single-bit tag per capability, but stores poison capabilities within the memory data itself. Moreover, poison capabilities contain substantially more metadata than
small external memory tags, allowing PoisonCap to encode the allocation bounds required to enforce temporal safety across multiple memory allocation layers.

\section{Threat Model}
PoisonCap embraces a more comprehensive threat model than Cornucopia Reloaded.
PoisonCap assumes that attackers might:
\begin{enumerate}
    \item exploit accesses through stale capabilities after the program calls \texttt{free()} but before reallocation;
    \item leverage heap read-before-write program errors to trigger undefined behavior.
    \item leverage either of the above violations in a nested allocator.
\end{enumerate}
While stack temporal safety is also important, we consider it as out-of-scope and future work.

\subsection{UAF Mitigation vs. UAR Mitigation}

PoisonCap embraces a strict Use-after-Free (UAF) threat model, rather than the more relaxed Use-after-Reallocation (UAR) threat model.
UAR mitigation
may allow stale capabilities to remain usable while storage is
quarantined, but ensures that they are revoked before the storage is reallocated~\cite{ainsworthMarkUsDropinUseafterfree2020, erdosMineSweeperCleanSweep2022, Cornucopia_reloaded}. This provides strong protection against attacks that rely on memory reuse, but does not enforce access denial between the program's call to \texttt{free()} and subsequent reallocation.

PoisonCap targets this additional temporal-safety property. Once the allocator processes \texttt{free()}, PoisonCap poisons the allocation and rejects subsequent accesses through capabilities belonging to it, even while its memory remains quarantined. Thus, PoisonCap does not challenge the safe-reclamation guarantee provided by deferred reclamation; it provides the stronger access denial property that stale capabilities cannot continue accessing an allocation after \texttt{free()}.

To conclusively address the enforcement gap between free and reallocation, PoisonCap therefore adopts the stronger policy that any access through a stale capability following free() is a temporal-safety violation and must be prevented.

\subsection{Use-After-Free Violations in Nested Allocators}
As the complexity of software stacks increase, UAF violations increasingly arise in nested memory allocators,
exposing layered software to security vulnerabilities.
Existing publications that address UAF problems have restricted enforcement to the libc memory allocation layer due to the difficulty of scaling UAF mitigation to multiple allocation layers~\cite{ainsworthMarkUsDropinUseafterfree2020,erdosMineSweeperCleanSweep2022,CHERIoT,ARM_MTE,sasakiPracticalByteGranularMemory2019,Cornucopia_reloaded}.
As listed in Table~\ref{table:nested_cwes}, managed programming languages and user-space applications that use internal custom allocators or caching layers to optimize performance (e.g., CPython, PostgresSQL, SQLite, OpenSSL, and Apache) would not benefit from single-layer UAF protection but are nevertheless included in the trusted computing base for many users.

Ideally, CHERI memory safety would nest elegantly to allow protection at every level.
CHERI spatial safety is able to address the nested buffer overflow and out-of-bounds reads listed in Table~\ref{table:nested_cwes} due to unprivileged instructions to subset the bounds of any capability.
However, it is challenging to extend Cornucopia Reloaded CHERI temporal safety to nested allocation layers in a scalable manner.

Use-after-free prevention across multiple allocation layers requires correctly distinguishing between legal and illegal accesses, as well as capabilities that need to be revoked and those that are still valid.
A program might hold dangling capabilities to a suballocation that has been freed; at the same time, the nested allocator will hold capabilities pointing to the slab, and the libc allocator will hold capabilities pointing to a parent slab.
Any nested UAF mechanism must prevent access to the dangling suballocation, while allowing access through the slab capabilities.
Equally, the sweeping revoker must distinguish dangling capabilities from broader slab capabilities.
Nested UAF protection should extend to any internal memory management, including structures that are not traditionally considered to be allocators.
For instance, UAF violations have arisen in typed allocation caches~\cite{ExtensionTypesCythona,CVE-2010-5298}, such as the freelist in OpenSSL~\cite{ CVE-2010-5298}.

Cornucopia Reloaded currently does not support temporal safety in nested allocation layers.
Cornucopia Reloaded relies on a single shadow bitmap provided by the kernel revocation service to identify memory freed in a single layer and uses a single software-defined permission bit to distinguish capabilities belonging to the allocator (which should escape revocation) from those belonging to the application.
This approach does not scale elegantly to more than one allocation layer, potentially requiring multiple shadow bitmaps and ``allocation level'' bits in each capability to direct the revoker to the relevant bitmap.

Our PoisonCap threat model includes UAF violations in nested allocators to allow full, nested temporal safety. PoisonCap therefore extends temporal-safety enforcement across nested
allocation layers.

\subsection{Uninitialized Access} 

Uninitialized memory access, leading to undefined behavior, remains a major security problem in modern software.
In particular, as reported by MSRC, 5 to 10\% of the Common Vulnerability Enumerations (CVEs) issued by Microsoft are uninitialized memory vulnerabilities~\cite{cheri_security_analysis,msrc_uninitialised}. 

\emph{Memory zeroing} is a common security strategy in programming languages and operating systems to prevent both information leakage and uninitialized memory access; indeed, memory zeroing is considered best practice for current CHERI systems.
However, MSRC found that zeroing is not sufficient to ensure initialization safety~\cite{msrc_uninitialised}.
Zeroing was found to permit buggy programs to silently continue execution with a predictable zero value rather than faulting, leading to a pointer that is not correctly initialized triggering a ``NULL pointer'' code branch~\cite{msrc_uninitialised}, leading to undefined behavior.

As shown in Table~\ref{table:nested_cwes}, the uninitialized access vulnerabilities found in OpenJPEG and libjxl cannot be mitigated by CHERI~\cite{CVE-2025-54874,CVE-2025-12474} or zero initialization.
These vulnerabilities are software logic failures, and zero initialization does not resolve these issues but rather obscures them, causing these programs to execute on a wrong path and produce incorrect output.

The PoisonCap threat model includes uninitialized access for heap allocations to ensure that CHERI software does not silently propagate undefined behavior.

\begin{figure*}[t]
    \centering
    \begin{subfigure}[t]{0.48\textwidth}
        \centering
        \def\svgwidth{\linewidth}
        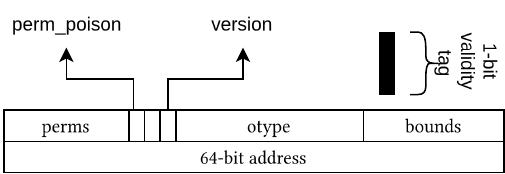
        \caption{CHERI capability extended with poison support.}
        \label{fig:memory_cap_format}
    \end{subfigure}
    \hfill
    \begin{subfigure}[t]{0.48\textwidth}
        \centering
        \def\svgwidth{\linewidth}
        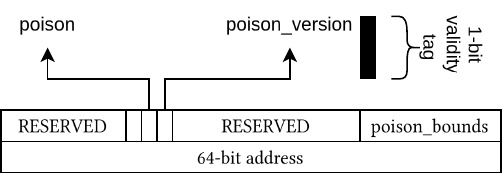
        \caption{Poison capability painted into freed memory.}
        \label{fig:poison_cap_format}
    \end{subfigure}
    \caption{(a) Capability format with poison support. (b) Poison capability written to freed memory.}
    \label{fig:cap_formats}
\end{figure*}

\section{PoisonCap Implementation}
PoisonCap aims to address the aforementioned threats by providing
hierarchical heap temporal-safety and initialization-safety support on CHERI.
PoisonCap relies on a new CHERI-RISC-V architectural extension, \emph{poison capabilities}, which track pointer/memory relationships across multiple allocation
layers without requiring additional external metadata or per-layer
capability metadata.
In this section, we present our poison capability instruction-set extension and how we have used it to construct scalable defenses against these classes of temporal safety vulnerabilities.

The poison capability architectural extension includes:
\begin{itemize}
    \item A \emph{poison bit} indicates capabilities that restrict access to the memory location where they are stored.
    \item A \emph{perm\_poison} permission allows a capability to either store poison capabilities into memory or to \emph{detox} memory by overwriting poison capabilities.
    \item A special store \emph{cpoison} instruction poisons memory.
    \item Superallocation management: reads and writes to poisoned memory are permitted when the bounds in the pointer are broader than the bounds of a poison capability that it references.  This ensures that a ``more powerful'' allocator (including the kernel) can freely manage memory that may have been poisoned by a downstream allocator.
    \item A \emph{version field} detects accesses to a freshly reused allocation~\ref{subsec:versioninig}, potentially trapping on read and auto-zeroing on write.
    \item A \emph{CGetPoison} instruction detects if a memory location is poisoned without causing a trap.
    \item A \emph{CGetCapPoison} instruction allows the kernel to determine if a location holds a poison capability supporting sweeping revocation and to allow marking pages as capability-clean, even if they contain poison capabilities. 
\end{itemize}

\subsection{Poison Capabilities}

Poison capabilities are special values stored in memory. The integrity of these values is protected using CHERI tagged memory; the special values are just special capability values. When any load or store touches this memory, the microarchitecture prevents it by either a trap or silent cancellation. 
Figure~\ref{fig:poison_cap_format} shows the format of a \emph{poison capability} that we have added to the CHERI-RISC-V instruction set architecture.
A poison capability is distinguished from a memory capability by a single \emph{poison} bit, which is set to 1 for poison capabilities.
Unlike typical CHERI capabilities,
poison capabilities do not grant access to memory, but prevent the memory location in which they are stored from being read or written.

Poison capabilities also have a \emph{poison version} field to allow references from new allocations, and preserve the bounds of the allocation that has been poisoned to allow accesses from more powerful allocations.
Unlike memory tagging designs such as ARM MTE which are limited in external tag size (\emph{e.g.} 4-bits associated with each 16 bytes of data)~\cite{ARM_MTE}, poison capabilities are stored in-line, allowing nearly 128-bits of tag metadata for each word of freed memory.
This allows poison capabilities to track more information about freed memory, including a 1-bit version and the level of revocation, providing more flexible enforcement than MTE.
As with ARM MTE, PoisonCap checks memory contents before allowing a memory operation to complete, but unlike MTE, it does not use tags in a shadow space, and instead relies on the existing tagged-capability mechanism to encode special values in memory itself.

Figure~\ref{fig:memory_cap_format} shows our changes to the format of a memory capability in our system.
We have added a memory version field, corresponding to the poison version field, which is used to allow new references to old allocations that have been poisoned and subsequently had all references revoked.
We have also added a \emph{perm\_poison} permission to allow this capability to be used to poison memory using the new poison instruction, to overwrite poisoned memory, and to update the memory version of a capability.

\subsection{Hardware Implementation}
\label{subsection:HW_impl}
We implemented PoisonCap in CHERI-Toooba, a superscalar, out-of-order CHERI RISC-V CPU~\cite{ruggSuiteProcessorsExplore2024}, adding new instructions, a PoisonCap format, and microarchitectural support for managing and detecting poisoned memory.

In CHERI-Toooba, memory accesses undergo capability validation, including tag, sealing, bounds, and permission checks, before address translation and cache access.
PoisonCap preserves this flow, and additionally adds circuits to check the 128-bit region returned by the cache, regardless of access size.
If the word has a valid CHERI tag and its \textit{POISON} bit is set, a load raises a PoisonCap exception before register writeback.
For stores, PoisonCap cancels poison violations without raising an exception.
This asynchronous behavior is similar in hardware complexity to ARM's Asymmetric mode in MTE, which traps on load violations, but not on store violations.
Asymmetric and Asynchronous MTE report a store failure in a pending fault register which is later checked by software, but PoisonCap is able to precisely cancel the store, as external metadata is not required for the poison check.

The PoisonCap implementation reuses CHERI-Toooba's existing capability validation circuits, capability decoding/decompression, cache pipeline, reorder buffer, and load-store queue. PoisonCap adds recognition of the PoisonCap encoding, poison/version comparison, and control logic in the L1 response path, and carries the authorizing capability's bounds information through the memory pipeline so that the L1 can perform the hierarchical poison-bounds check described in Section~\ref{subsection:nested_revoke}.

PoisonCap performs poison validation at the L1 cache, where the previous value of the data is naturally available.
As the L1 cache is after the store queue, this opens a corner case for store-to-load forwarding: a younger load may receive the value of an unchecked store directly from the store queue before the store reaches the L1 poison check.
If the store targets poisoned memory, the L1 check subsequently cancels the store, but the younger load may already have consumed the forwarded value.
This allows a violation of use-after-free for tight write-read sequences.
Asymmetric MTE implementations similarly suffer from store-queue forwarding between violating stores and good loads~\cite{kim2025tiktag}.
Asymmetric MTE will nevertheless note the store violation in the pending fault register; PoisonCap in CHERI-Toooba will silently cancel the store and leave no indication of failure. 
A non-prototype specification for PoisonCap should consider options to either indicate canceled stores, or strictly check all loads from the cache, even if speculatively forwarding from the store queue.

\subsection{Poison Bounds for Revoking Nested Dangling Capabilities}
\label{subsection:nested_revoke}
PoisonCap exploits the \emph{monotonicity} of CHERI capabilities to distinguish allocation layers without external metadata. Because CHERI capability bounds can only be monotonically decreased, deriving a capability for a sub-allocation narrows its bounds, and these bounds cannot subsequently be widened. When the sub-allocation is freed, PoisonCap records these bounds as \emph{\texttt{poison\_bounds}}. Consequently, comparing the decompressed bounds of a capability with the stored \texttt{poison\_bounds} identifies capabilities referring to the freed allocation while preserving broader capabilities owned by upstream allocators.

Listing~\ref{lst:nested} and Figure~\ref{fig:poison-memory} illustrate this process using a parent allocator that provides a memory region to a child allocator. Initially, the child allocator creates two sub-allocations, \texttt{x} and \texttt{y}. When \texttt{x} is freed, the child allocator overwrites the object with PoisonCaps containing \texttt{poison\_bounds} equal to the bounds of \texttt{x}. Unlike a NULL capability, PoisonCap preserves the metadata required to identify dangling capabilities to a sub-allocation and does so without relying on external shadow metadata.

\begin{algorithm}
\caption{\textsc{MatchPoison}}
\label{alg:matchpoison}
\DontPrintSemicolon

\KwIn{Capability $\mathrm{cap}$, PoisonCap $\mathrm{pcap}$}

\Return{
$\mathrm{pcap}.\mathrm{tag} \land \mathrm{pcap}.\mathrm{poison} \land \neg \mathrm{range}(\mathrm{cap}).\mathrm{supersetOf}(\mathrm{pcap})$
}
\end{algorithm}

\begin{lstlisting}[
    language=C,
    caption={Nested allocation and poisoning. Labels (A), (B), and (C)
    correspond to the states in Figure~\ref{fig:poison-memory}.},
    label={lst:nested}
]
void *p = malloc(64);               // Parent allocation

{
    void *c = parent_alloc(p, 48);  // Create child

    void *x = child_alloc(c, 16);
    void *y = child_alloc(c, 16);   // (A)
    
    child_free(x);                  // (B)
    
    // x is dangling, *x filled with [bx, tx] poison
    *x = 42;                        // Cancelled
    printf("%d\n", *x);             // PoisonFault

    // c has wider bounds, so can re-init x when needed
    memset(c, 16, 0);               // OK

    parent_free(c);                 // (C)
    // *c filled with [bc, tc] poison, no longer usable
}

// p has wider bounds, so can re-init
memset(p, 64, 0);                   // OK
\end{lstlisting}

Algorithm~\ref{alg:matchpoison} is used by hardware when detecting
a poison-bounds violation to enable hierarchical revocation.
Algorithm~\ref{alg:matchpoison} compares the decompressed bounds of a dereferenced capability with the \texttt{poison\_bounds} stored in a PoisonCap. There is a match if the pointer bounds are not a superset of the \texttt{poison\_bounds}, indicating that the pointer capability originated from the freed allocation. 
Consequently, matching loads raise a \textsc{PoisonException}, matching stores are canceled, and matching capabilities encountered during revocation have their tags cleared.
More powerful capabilities belonging to upstream allocators have wider bounds and therefore do not match the stored \texttt{poison\_bounds}.

\begin{figure}[h]
\centering
\footnotesize
\setlength{\tabcolsep}{5pt}
\renewcommand{\arraystretch}{1.2}

\textbf{(A) Before \texttt{child\_free(x)}}\\[1mm]

\begin{tabular}{|c|c|c|c|}
\multicolumn{4}{c}{memory map} \\[0.4em]
\multicolumn{4}{c}{$b_p$\hspace*{\fill}Parent region (p)\hspace*{\fill}$t_p$} \\[0.4em]
\multicolumn{3}{c}{$b_c$\hspace*{\fill}Child region (c)\hspace*{\fill}$t_c$} & \multicolumn{1}{c}{} \\[0.4em]
\hline
\rule{0pt}{1.25em} && \multirow{4}{*}{\rotatebox[origin=c]{90}{\parbox{1.25cm}{\centering c metadata}}} &  \multirow{4}{*}{\rotatebox[origin=c]{90}{\parbox{1.25cm}{\centering p metadata}}} \\
\parbox{1.2cm}{\centering{\textbf{x}}} & \parbox{1.2cm}{\centering{\textbf{y}}} & & \\
 & & & \\
$b_x$ \rule{0pt}{1.25em} \hspace*{\fill} $t_x$ & $b_y$ \hspace*{\fill} $t_y$ && \\
\hline
\end{tabular}

\vspace{2mm}

The parent allocator controls the parent region.\\
The child allocator controls the child sub-region,\\
and has allocated two sub-sub-regions x and y.

\vspace{2mm}

\textbf{(B) After \texttt{child\_free(x)}}\\[2mm]

\begin{tabular}{|c|c|c|c|}
\hline
\rule{0pt}{1.25em} && \multirow{4}{*}{\rotatebox[origin=c]{90}{\parbox{1.25cm}{\centering c metadata}}} &  \multirow{4}{*}{\rotatebox[origin=c]{90}{\parbox{1.25cm}{\centering p metadata}}} \\
\parbox{1.2cm}{\centering\textbf{POISONED}} & \parbox{1.2cm}{\centering\textbf{y}} & & \\
$[b_x,t_x]$ & & & \\
\rule{0pt}{1.25em} &&& \\
\hline
\end{tabular}

\vspace{2mm}

x cannot be accessed through dangling references, but can be\\accessed by a larger parent region or child region capability.

\vspace{2mm}

\textbf{(C) After \texttt{parent\_free(c)}}\\[2mm]

\begin{tabular}{|c|c|c|c|}
\hline
\multicolumn{3}{|c|}{\rule{0pt}{1.25em} } & \multirow{4}{*}{\rotatebox[origin=c]{90}{\parbox{1.25cm}{\centering p metadata}}} \\
\multicolumn{3}{|c|}{\parbox{3.25cm}{\centering\textbf{POISONED}}} & \\
\multicolumn{3}{|c|}{\parbox{3.25cm}{\centering$[b_c,t_c]$}} & \\
\multicolumn{3}{|c|}{\rule{0pt}{1.25em} } & \\
\hline
\end{tabular}

\vspace{2mm}

The child region cannot be accessed by the child,\\but can be accessed by a larger parent region capability.



\caption{
Memory state of a nested allocator as memory is allocated and freed.
Freed memory is filled with PoisonCaps encoding the
\texttt{poison\_bounds} of the freed range. 
During memory accesses and revocation, capabilities that do not have bounds broad enough to contain the stored \texttt{poison\_bounds} are denied or revoked, while broader capabilities held by upstream allocators remain valid.
}
\label{fig:poison-memory}
\end{figure}

\subsection{Revocation Without a Shadow Bitmap}
\label{subsec:noshadow}
\begin{figure}[h]
        \centering
        \def\svgwidth{\linewidth}
        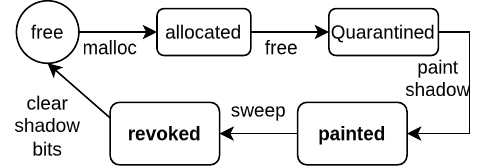
        \caption{Cornucopia Reloaded CHERI heap memory lifetime~\cite{Cornucopia_reloaded}}
        \label{fig:cornu_flow}
\end{figure}

As described in Section~\ref{subsec:Cornucopia}, Cornucopia Reloaded uses a software-managed shadow bitmap to record allocations awaiting reclamation. 
During a revocation sweep, the revoker consults this bitmap to determine whether each capability refers to such an allocation. PoisonCap retains Cornucopia's quarantine-and-sweep structure, but replaces this lookup with the poison state stored directly in the allocation.

We modify the Cornucopia revoker so that, instead of consulting the shadow bitmap, it probes memory at the base address of each capability under test. If a PoisonCap is found, the capability bounds are compared with the stored \texttt{poison\_bounds} using the rule described in Section~\ref{subsection:nested_revoke}. 
Capabilities belonging to the poisoned allocation are revoked, while broader capabilities held by upstream allocators remain valid. 
Thus, the PoisonCap written when the allocation is poisoned serves both to restrict access to the allocation and to provide the metadata required by the subsequent revocation sweep.
Because the allocation layer is encoded by \texttt{poison\_bounds}, the same mechanism supports nested allocators without maintaining a separate shadow bitmap for each allocation layer.

The revoker must be able to inspect poisoned memory without triggering PoisonCap's normal access restrictions. 
It therefore re-derives the capability used for the probe from the userspace root capability, which bears the \emph{perm\_poison} permission. The capability under test may also be sealed, have an unaligned base, or have bounds smaller than a capability word. 
Re-derivation together with appropriate alignment handling allows the revoker to probe the relevant PoisonCap-sized region without causing a CHERI protection violation. 
For correctness, this operation assumes that the user allocation granule is at least the size of a capability; otherwise, the aligned probe could access memory belonging to a different allocation.

A further complication is that of `recursive page faults'.
The Cornucopia revoker operates on a page-by-page basis: when a new page is pulled in via a page fault, either by a revocation sweep or directly by the application, it is first scanned by the revoker to invalidate old capabilities before being passed to the runtime.
During a page visit, the Cornucopia revoker may page fault a second time when accessing the shadow bitmap.
These recursive page faults are easily handled because the shadow mapping can never store capabilities, so the page that gets faulted-in does not need to be recursively scanned.
Conversely, the PoisonCap revoker directly dereferences capabilities within the visited page.
If dereferencing the capability under test creates a second page fault, the new page may itself hold capabilities and would need to be recursively scanned before the page is mapped in and exposed to the application.

Our prototype assumes the system does not page-out memory and the program does not have aliased mappings that may store capabilities, including CoW (Copy on Write) mappings.
This ensures that no new capability-holding memory will need to be paged-in, avoiding the problem of recursive page faults and scans.
While these assumptions are fairly restrictive, the goal of this prototype is to demonstrate the principle of operation of the revoker.
Furthermore, we still support many beneficial parts of the virtual memory system, such as zero-filled anonymous pages and memory-mapped files, because their pages cannot hold capabilities.

This implementation shows that it is feasible to modify the revoker to operate without a shadow bitmap, and to support nested allocators; however, the current prototype requires further engineering work to be equally robust to the Cornucopia revoker. 
We believe it should be possible for the PoisonCap revoker to handle recursive faults, by temporarily mapping the second page in with limited permissions and deferring the recursive scan, but we leave this and performance optimization as future work.

\subsection{Nested Temporal Safety for SQLite}
To understand and evaluate the feasibility of PoisonCap support for nested temporal safety in real-world applications, we have extended the nested allocator in SQLite to use PoisonCap for temporal safety.
SQLite supports multiple options for built-in custom allocators which obtains heap memory from the libc allocator through malloc() and manages internal memory allocations without interacting with the parent allocator.
We chose the MEMSYS5 allocator of SQLite for a proof-of-concept PoisonCap nested allocator.
For spatial safety, we extended MEMSYS5 with native CHERI support.
For temporal memory safety, we extended the MEMSYS5 allocator to poison memory on free to support deterministic strict use-after-free mitigation while also allowing the revoker to identify dangling references and maintain a quarantine buffer to delay reallocation, as per Cornucopia Reloaded.

Support for poison bounds allows MEMSYS5 to safely free sub-allocations without compromising libc slab capabilities or its own arena capabilities.
As the poison bounds written to freed memory are subsets of the root capability in the libc allocator and the arena capability maintained by MEMSYS5 allocator, the revoker can revoke dangling capabilities within the MEMSYS5 allocator without revoking the capability maintained by the libc allocator and MEMSYS5's arena allocator.

In addition, our modified MEMSYS5 can poison and detox memory without relying on the \emph{perm\_poison} permission, which is cleared on capabilities returned from the libc allocator, as the broader bounds of its arena capabilities authorize it to perform these privileged functions. This implementation proved out the nestability of PoisonCap using poison bounds, without relying on static allocation layers implied by the \emph{perm\_poison} permission.

\subsection{Versioning for Initialization Safety}
\label{subsec:versioninig}
Initialization safety can be efficiently enforced using a write-before-read policy; that is, words of a new allocation must be written before they are read, ensuring that program behavior cannot be affected by undefined memory values.
In order to detect accesses to uninitialized allocations, we define 1 bit of capability metadata to be a \emph{memory version}, as shown in Figure~\ref{fig:poison_cap_format} and Figure~\ref{fig:memory_cap_format}. 
Every capability returned on allocation is assigned the opposite version of its predecessor.
Freed memory is poisoned with capabilities of the same version as the freed allocation.

Use-after-free is prevented by trapping if a pointer capability version matches the poison capability version found in the memory it attempts to access.
If these versions match, the memory access traps to prevent use-after-free.  Uninitialized access is prevented by trapping on a read through a pointer capability version that is different from the version of a poison capability in the memory being accessed.
Writes through capabilities of a version different from the poison they overwrite are allowed.
Narrow writes will zero the remainder of the capability they access, trading byte-granularity of write-before-read enforcement for a practical metadata granularity.

Tracking the memory version in the allocator can be difficult.
Jemalloc, the CheriBSD system allocator, only holds pointers for allocations stored in its fast path, thread-local cache (TCache)~\cite{Jemalloc}; this naturally preserves the previous memory version for future allocations.
In other cases, we must read the memory version from the poison capabilities that have been written to the freed allocation.

Reusing a capability \texttt{memory\_version} for different allocations is safe because revocation occurs between version changes. 
Before a version is reused, the sweep revokes surviving capabilities whose version matches the poisoned allocation and whose bounds fail the poison-bounds check described in Section~\ref{subsection:nested_revoke}.
Thus, after the sweep, no stale capability that would be denied access to that poisoned allocation survives into the next reuse of the version.

PoisonCap currently supports only quadword-granularity (128-bit) initialization safety; a legal write that is smaller than a capability-word detoxes the entire capability word of poison and auto-zeros the memory.
That is, the non-written bytes of the capability word fall back to zero initialization and would not trap on read.

\subsection{Using Poison to Improve Cache Efficiency}

In previous evaluations of Cornucopia, the CPU's memory subsystem has not been aware of memory quarantining, resulting in caches evicting useful data and holding onto lines that are not expected to be accessed again.
Zeroing memory does not resolve this problem, as a zero in a cache line is not a reliable indicator of whether a cache line is in use; indeed, zeroing is often used to prefetch lines into the cache on allocation.
In contrast, poisoning memory clearly communicates the software semantics of freed memory to the memory subsystem.
As shown in Figure~\ref{fig:poison_cap_format}, poison capabilities are easily distinguishable in memory.
The memory system can simply read two bits -- the POISON bit and the CHERI tag -- from each 16-byte word to determine if the 16-byte word is a poisoned capability.
When a memory store updates a cache line in the L1 data cache, the memory subsystem marks if the updated cache line exclusively contains poison capabilities. 
We tweaked the PoisonCap CHERI-Toooba cache replacement algorithm to prioritize replacing poisoned cache lines, reserving cache space for memory that is not freed.

We should note that certain vendor-provided zeroing instructions, such as ARM ``dczva'' and RISC-V ``zbo.zero'', are extended to be cache locality aware to optimize cache efficiency of zeroing.
In addition, RISC-V has a ``Zihintntl'' extension that provides hints that a subsequent load or store instruction will not exhibit temporal locality \cite{81ZihintntlExtension}, which can also be used to hint that zeroed lines will not be used to reduce cache pollution.
Either due to architecture or common microarchitectures, these mechanisms work on a cache-line granularity, and would not allow the caches to aggregate contiguous blocks of smaller freed allocations.
PoisonCap allows caches to detect poisoned memory at a word granularity and naturally aggregates poisoned allocations into full poisoned cache lines to be marked for replacement.
PoisonCap folds efficient cache management of freed memory into a single operation on free which both marks freed memory in the caches and auto-zeros on reallocation, as described in the next section.

\section{Evaluation}
To evaluate PoisonCap security, performance, and hardware efficiency, we use the following methods and benchmarks:
\begin{itemize}
    \item To evaluate PoisonCap's temporal safety enforcement, we use version 1.3 of the U.S. NIST SARD Juliet Test Suite~\cite{Juliet} and MSET benchmark suite~\cite{vintila2025mset}.
    \item To evaluate whether PoisonCap can scale temporal safety to applications' custom nested allocators, we evaluated PoisonCap on SQLite which has a suite of custom allocators. 
    \item To evaluate performance, we ran SQLite, SPEC CPU2006 INT and PARSEC benchmarks on PoisonCap extended CHERI-Toooba on the VCU118 FPGA platform. 
    \item To evaluate the hardware implementation overhead of PoisonCap on CHERI-Toooba, we used the area and utilization reports generated by Vivado 2018.
\end{itemize}

\subsection{Experimental Setup}
PoisonCap is a hardware-software co-design whose implementation spans the CHERI software stack, including the LLVM compiler toolchain, the CheriBSD operating system, and Jemalloc user-space allocators, to processor implementations including both the QEMU emulator and CHERI-Toooba.
We use PoisonCap's QEMU implementation to verify its architectural behavioral correctness and to evaluate its compatibility with the CHERI software stack.
For performance and hardware impact evaluations, we extend CHERI-Toooba with PoisonCap running on our VCU118 based FPGA platform.

CHERI-Toooba on our VCU118 FPGA platform is configured to have 2 CPU cores.
Each of the cores has a 4-way, 32 KB L1 data cache, and the 2 cores share a 16-way 1 MB last-level cache that is connected to the CHERI tag controller and DRAM.
Our PoisonCap extension of CHERI-Toooba is synthesized for the VCU118 FPGA using Vivado 2018 at a 25 MHz clock frequency.

\subsection{Area and Power Impact on CHERI-Toooba}

The PoisonCap implementation in CHERI-Toooba introduced only minimal overhead.
PoisonCap is designed to avoid complex changes to timing-critical and memory access paths so as not to affect common-case memory access performance, timing closure, and power efficiency.
PoisonCap's major hardware circuit changes include instruction decoding support for poison instructions, and poison detection units added for both memory load and store.
PoisonCap also adds a 1-bit tag in the caches to mark fully poisoned cache lines and uses this tag to guide the cache replacement algorithm in both the L1 data cache and the L2, last-level cache in CHERI-Toooba.
PoisonCap adds 1.1\% to logic usage and 2.2\% register usage to CHERI-Toooba synthesized for the UltraScale+ FPGA on the VCU118.
Passing the 64-bit capability length through the memory pipeline to compare against loaded poison bounds contributes to a large proportion of register increases.
As future work, this cost could be reduced by representing the length using compression similar to CHERI Concentrate~\cite{woodruffCHERIConcentratePractical2019}.

\subsection{CHERI Software Stack Compatibility}
To develop and verify the feasibility of PoisonCap use-after-free enforcement in software, we enabled freed-memory poisoning across the entire CHERI software stack, including CheriBSD and user-space libraries. Due to the complexity of the software stack, supporting poison capabilities across the entire software stack correctly and securely is a challenge.
We have learned a few lessons from bringing up the CHERI software stack on PoisonCap hardware: 

\textbf{Memory poisoning and poison-clearing (or \emph{detox}) operations are not allowed to be executed arbitrarily by unauthorized entities.}
For use-after-free mitigation in the libc allocation layer, user-space application code must not be allowed to detox memory (to avoid use-after-free); these operations must be restricted to the kernel and the libc allocators.
To enforce this property, we require writes to poisoned memory to be authorized by a capability with either a different version or broader bounds compared to the poison capability in memory. \textbf{Allocators must also be able to detox memory after revocation before reuse}, and they must hold capabilities of broader bounds to ensure that they do not trap or have their capabilities revoked.

\textbf{The kernel needs to write to poisoned memory}.
PoisonCap forbids application code from overwriting poisoned memory, but we found that the kernel must be allowed to write to poisoned memory in order to zero reclaimed physical pages for reuse, which may contain poisoned memory.
If poison in a page is not zeroed by the kernel, previously poisoned memory can be remapped to another application, causing it to fail; even the init process of CheriBSD fails during boot. 
    \textbf{The kernel needs to copy poisoned memory (\emph{i.e.}, both read and write)}.
To support copy-on-write for virtual memory, the kernel needs to copy potentially poisoned memory contents from one page to another.
To permit the kernel to write to and read from poisoned memory, we enabled the \emph{perm\_poison} permission for all kernel capabilities.
    
\textbf{The revoker needs to identify whether a location is poisoned} but does not need to load the contents of the capability.
To support this use-case, we added a \emph{CGetPoison} instruction that returns a single bit indicating the poison state of a memory location, allowing the revoker to skip examination of these words. \textbf{The kernel must also treat poison capabilities as non-capability} data in order to mark a page with freed memory as \emph{clean}, that is, devoid of valid capabilities; we use \emph{CGetCapPoison} for this as well.

We also experimentally extended the CHERI capability format with a \emph{perm\_poison} permission to always authorize poison and detox, regardless of bounds, and also to change the memory/poison version for initialization safety.
This mechanism is not natively hierarchical; hierarchical initialization safety in PoisonCap would require a privileged software service to update the capability version for nested allocators.

\subsection{Security}
\label{subsubsec:hierarchical_eval}
\begin{figure*}[t]
    \centering
    \includegraphics[width=\textwidth]{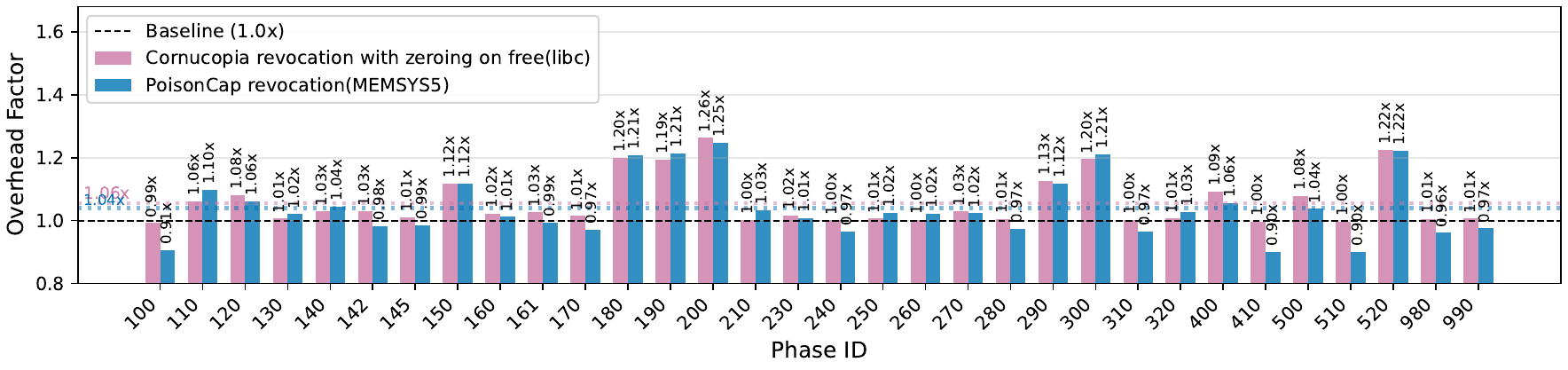}
     \caption{Normalized Performance overhead of Cornucopia Revocation overhead on SQLite that uses libc allocator and PoisonCap overhead on SQLite that uses nested SQLite MEMSYS5 allocator across SQLite speedtest1 operational phases.}
    \label{fig:sqlite_perf}
\end{figure*}
For security evaluation, we use version 1.3 of the U.S.
NIST SARD \textbf{Juliet Test Suite}~\cite{Juliet,Juliet_test_doc} which contains thousands of C/C++ related memory vulnerabilities.
In order to evaluate PoisonCap's effectiveness at mitigating temporal safety violations, we tested against the following relevant classes: CWE-415: Double Free, CWE-416: Use After Free, and CWE-457: Uninitialized access. We additionally evaluate PoisonCap for use-after-free mitigation using \textbf{MSET}~\cite{vintila2025mset}, a recently proposed framework for systematically generating memory-safety test programs. Compared with the manually written Juliet suite, MSET provides broader coverage of temporal memory-safety patterns but does not include initialization-safety tests. Therefore, MSET is used to complement the Juliet Test Suite's evaluation of temporal safety.

\subsubsection*{Use-after-free}
\label{uaf_eval}
The default configuration of CHERI Cornucopia temporal safety fails to detect use-after-free violations that occur between free and reallocation in the CWE-416 class from \textbf{Juliet Test Suites}.
A debug mode of Cornucopia can, in fact, detect these cases at a crippling cost to performance by configuring a zero-size quarantine buffer so that it revokes on every free. PoisonCap is able to run all the cases provided for CWE-416 and mitigates all vulnerabilities in the ``bad'' cases. 

We also evaluated PoisonCap using all 12 generated heap use-after-free tests from the \textbf{MSET} benchmark suite~\cite{vintila2025mset}. The outcome of each test is summarized in Table~\ref{tab:mset-uaf-results}. During execution, 4 tests were automatically skipped by the MSET framework because their runtime preconditions were not satisfied. Of the remaining 8 tests, 4 triggered conventional CHERI invalid tag exceptions because the stale capability was dereferenced after the freed object had been reallocated, allowing the revocation sweep to revoke the stale capability before the access. The remaining 4 tests dereferenced stale capabilities after free but before reallocation, triggering the PoisonCap exception. These results demonstrate that PoisonCap extends the temporal safety guarantees of Cornucopia Reloaded by detecting strict use-after-free accesses while preserving protection against use-after-reallocation attacks. Across all executable MSET tests, PoisonCap successfully detected temporal memory safety violations arising from both direct loads/stores and standard library memory operations.

\begin{table*}
\centering
\caption{Outcomes of the MSET use-after-free tests. }
\label{tab:mset-uaf-results}
\small
\begin{tabular}{p{0.50\textwidth}c p{0.34\textwidth}}
\toprule
\textbf{Test} & \textbf{Outcome} & \textbf{Explanation} \\
\midrule

\texttt{use\_after\_star\_freed\_memory\_heap\_\{direct,stdlib\}\_\{read,write\}\_0}
&
POISON fault
&
The stale pointer accesses freed memory before reallocation, detected by PoisonCap precisely. \\

\texttt{use\_after\_star\_used\_memory\_heap\_\{direct,stdlib\}\_\{read,write\}\_0}
&
Skipped
&
Exited with status~43 before reaching the intended use-after-free access. \\

\texttt{use\_after\_star\_used\_memory\_heap\_\{direct,stdlib\}\_\{read,write\}\_1}
&
Tag fault
&
The access occurs after reallocation. Cornucopia revocation invalidates the stale capability, resulting in a conventional CHERI capability fault. \\

\bottomrule
\end{tabular}
\end{table*}

\subsubsection*{Initialization safety}
To evaluate PoisonCap's effectiveness for heap initialization safety, we focused on the CWE-457 (Use-of-Uninitialized variable) C test class, which has 176 ``good'' cases and an equal number of ``bad'' cases that are relevant to heap initialization safety.
PoisonCap successfully ran all good cases and detected all vulnerabilities in bad cases with no false positives.

In our survey of relevant vulnerabilities, we found that PoisonCap's write-before-read policy would detect and trap the uninitialized access vulnerability in OpenJPEG and libjxl~\cite{CVE-2025-54874,CVE-2025-12474} (listed in Table~\ref{table:nested_cwes}), resulting in efficient detection of the vulnerability rather than continued execution with undefined behavior.

PoisonCap's use-after-free protection is fully compatible with the current CheriBSD software stack.
The optional strict initialization-safety extension is not yet fully compatible with the current CheriBSD software stack. Enabling the extension exposes initialization-safety violations in existing user-space software, including copying uninitialized portions of dynamically allocated buffers (e.g., copying unused buffer capacity with `memcpy`) and reading partially initialized objects. Although these behaviors are often tolerated by conventional systems, strict initialization checking traps on such accesses. Addressing these compatibility issues will require targeted software, compiler, and library adaptions. We leave a systematic investigation of these cases, and the necessary software and compiler improvements, as future work.

\subsubsection*{Hierarchical use-after-free}
To prove the utility of PoisonCap for nested allocations, we extended SQLite's MEMSYS5 custom allocator with CHERI bounds for spatial safety and PoisonCap for temporal safety.
Extending MEMSYS5 with CHERI and PoisonCap support did not require structural software changes; no extra software defined permission was required to protect MEMSYS5's heap capabilities, and a shadow bitmap was not required for revocation at this layer.
Besides bounds for CHERI spatial safety and freed memory quarantining, PoisonCap required only poisoning on free and detox on reallocation.
The poison bounds primitive allows elegant delegation of poisoning and detox privileges to the MEMSYS5 allocator without disrupting the upstream allocator, allowing the revoker to revoke all capabilities that are downstream from the broadest capability to the memory that has been freed. 
During the integration of PoisonCap with MEMSYS5, we discovered that MEMSYS5 metadata was being stored in freed memory to track available allocations; this strategy is not safe without temporal safety.
To allow for poisoning freed allocations to enforce temporal safety, MEMSYS5's metadata is moved into an external memory region.

We found that PoisonCap's hierarchical use-after-free mitigation for nested allocations would have mitigated the nested use-after-free vulnerabilities~\cite{gBugHeapUse2024, CVE-2020-9273,cve-2019-0196}, as listed in Table~\ref{table:nested_cwes}.

We evaluated this extension using SQLite's built-in speedtest1 benchmark, which consists of multiple phases of database operations such as create, insert, reorder, and delete. 
The performance overhead for each phase of speedtest1 is given in Figure~\ref{fig:sqlite_perf} in the following configurations:
Baseline without Cornucopia Reloaded revocation; SQLite with Cornucopia Reloaded revocation;  SQLite with the PoisonCap-enabled MEMSYS5 allocator with POISON revoker. 
We were able to observe the correct revocation of capabilities allocated by MEMSYS5 and referencing poisoned memory without the use of a shadow bitmap or software defined permission to protect allocator capabilities.

This demonstrates that poison-based revocation can correctly identify and revoke dangling capabilities arising from a nested allocator in a real-world application.

\subsubsection*{Double free}
PoisonCap successfully passed all 1636 tests in the CWE-415: Double Free class from the \textbf{Juliet Test Suite}~\cite{Juliet}.
Unlike the default CHERI configuration, which relies on the shadow bitmap to detect double-free by checking if the shadow region has been painted prior to committing the paint, PoisonCap detects heap double-free by checking if the freed memory is already poisoned on memory free using the \emph{cgetpoison} instruction.

\subsection{Performance}

\subsubsection*{SPEC CPU2006 INT}
\label{subsubsec:spec_perf}
\begin{figure*}[htbp]
  \begin{subfigure}[b]{0.47\textwidth}
    \def\svgwidth{\linewidth}
    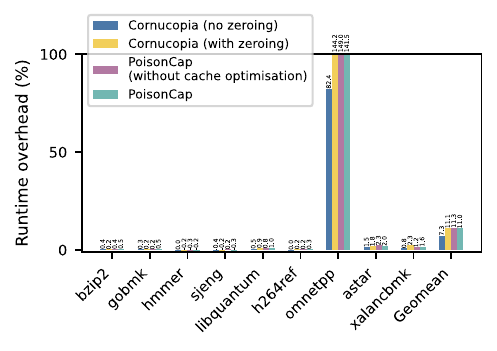
    \caption{Runtime overhead}
    \label{fig:cpu_cycle}
  \end{subfigure}
  \hfill
  \begin{subfigure}[b]{0.47\textwidth}
\def\svgwidth{\linewidth}
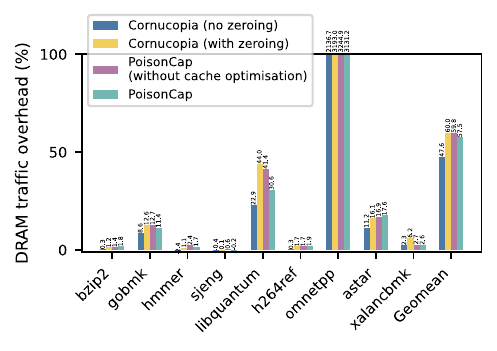
    \caption{DRAM traffic overhead}
    \label{fig:llc_miss}
  \end{subfigure}
  \caption{SPEC CPU2006 INT Overheads of Cornucopia Reloaded and PoisonCap relative to base CHERI without revocation.}
\end{figure*}
PoisonCap aims to enforce use-after-free protection without adding performance or memory overhead.
We use the CHERI-supported subset of SPEC CPU2006 INT benchmarks~\cite{henningSPECCPU2006Benchmark2006} to evaluate PoisonCap's impact on performance and memory usage.
We use the \emph{train} input dataset, which requires around 48 hours to complete a full run of 9 SPEC benchmarks on FPGA.
These benchmarks use the libc allocator, and we are using the Cornucopia kernel revoker that relies on a shadow bitmap.

We collect performance results with 4 temporal safety configurations:
CHERI without revocation; Cornucopia~\cite{Cornucopia_reloaded} revocation without zeroing; Cornucopia revocation with zeroing; and Cornucopia revocation with PoisonCap, with revocation disabled as the baseline.
While we include performance results without zeroing, we use Cornucopia with memory zeroing as the security-oriented performance baseline, since zeroing can prevent stale capabilities from observing previous allocation contents and provides predictable contents on subsequent allocations.
As shown in Figure \ref{fig:cpu_cycle}, PoisonCap does not introduce additional performance overhead relative to Cornucopia with zeroing.

For this performance evaluation, PoisonCap is integrated with the existing Cornucopia revoker, which continues to use its shadow bitmap. 
The purpose of this experiment is to measure the performance impact of replacing memory zeroing with PoisonCap while holding the revocation mechanism constant; nested revocation is evaluated separately.
Consequently, these SPEC results do not measure the potential performance or memory savings from eliminating the shadow bitmap, which requires a more mature poison-aware kernel revoker.

In fact, PoisonCap slightly improves performance on SPEC compared with Cornucopia with zeroing. 
In particular, it improves the overheads of allocation-intensive benchmarks such as omnetpp and xalancbmk. 
In particular, omnetpp performs over 130 million allocations and deallocations in the \emph{train} workload and exhibits the highest temporal safety overhead among the evaluated benchmarks. 
Consequently, Cornucopia Reloaded performs more than 2,000 revocation sweep cycles when executing omnetpp, while the high deallocation rate also triggers extensive memory zeroing, substantially increasing execution time. 
Although PoisonCap similarly overwrites freed memory, it writes poison capabilities instead of zeroes and benefits from the poison-aware cache management policy, reducing cache pollution and improving cache efficiency. 
Nevertheless, as shown in Figure~\ref{fig:llc_miss}, PoisonCap reduces overall DRAM traffic overhead, lowering the DRAM traffic overhead by an average of 1.49\%.

PoisonCap improves the cache efficiency of CHERI by managing freed memory flows at the microarchitecture level.
Improving last-level cache efficiency reduces DRAM traffic overhead and, consequently, memory-system power consumption. 
As shown in Figure \ref{fig:llc_miss}, the measured performance improvements are at least partially attributable to lower DRAM access costs enabled by the freed-memory-aware cache replacement policy. 
To isolate the contribution of this optimization, we evaluate PoisonCap both with and without the freed-memory-aware cache replacement policy. The results show that enabling poisoned cache line replacement reduces DRAM traffic overhead by a geometric mean of 2.3\% and lowers the overhead for 6 of the 9 SPEC benchmarks evaluated.
However, these benefits are offset by the reduced efficiency of CHERI's tag compression mechanism. The CHERI tag controller exploits the sparsity of tagged capabilities to compress tag metadata efficiently~\cite{joannouEfficientTaggedMemory2017,joannouHighperformanceMemorySafety2017}. In contrast to zeroing freed memory, PoisonCap overwrites freed objects with valid tagged poison capabilities, increasing the density of tagged capabilities. This reduces the effectiveness of tag compression, resulting in more tag-controller misses and a small performance penalty, as observed for benchmarks such as bzip2, libquantum, and h264ref.
PoisonCap also slightly increases revocation sweeping overhead. Because poison capabilities remain valid tagged capabilities, the revoker must load and inspect more tagged capabilities when searching for dangling capabilities, increasing the amount of work performed during the revocation pass.

\subsubsection*{PARSEC}
\begin{figure*}[t]
    \centering
    \begin{subfigure}[b]{0.47\textwidth}
        \centering
        \includegraphics[width=\linewidth]{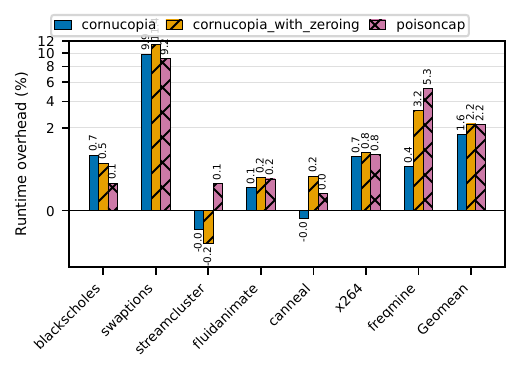}
        \caption{Single-thread runtime overhead}
        \label{fig:parsec-overhead-t1}
    \end{subfigure}
    \hfill
    \begin{subfigure}[b]{0.47\textwidth}
        \centering
        \includegraphics[width=\linewidth]{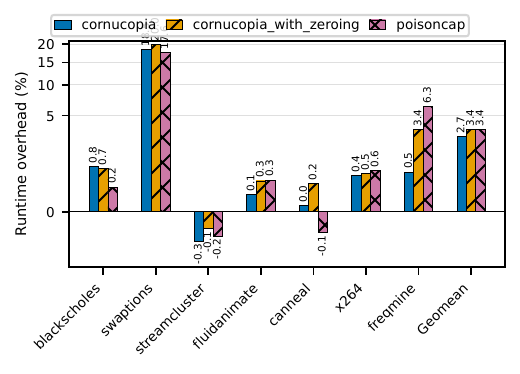}
        \caption{Two-thread runtime overhead}
        \label{fig:parsec-overhead-t2}
    \end{subfigure}
    \caption{PARSEC runtime overhead of Cornucopia Reloaded and PoisonCap relative to the baseline CHERI implementation without revocation.}
    \label{fig:parsec-overhead}
\end{figure*}

To further evaluate whether PoisonCap scales effectively on multi-threaded applications, we ported a representative subset of the PARSEC benchmark suite~\cite{bienia2008parsec}, comprising \textit{blackscholes, swaptions, streamcluster, fluidanimate, canneal, x264}, and \textit{freqmine}. These benchmarks were selected because they represent a diverse range of shared-memory workloads commonly used to evaluate multicore systems, complementing the single-threaded evaluation provided by SPEC CPU 2006 INT. Our FPGA evaluation platform, based on the Xilinx VCU118, is resource-constrained and can synthesize at most two CHERI-Toooba cores. As each core supports a single hardware thread, the PARSEC evaluation was limited to one- and two-thread executions to assess multicore scaling. Since PARSEC predates the CHERI purecap programming model, several benchmarks required source-level compatibility fixes to build and execute correctly under the CHERI-RISC-V purecap ABI. These changes addressed capability-safe pointer handling, legacy C/C++ compatibility issues, duplicate symbol definitions, and modern compiler compatibility while preserving the original benchmark algorithms. Seven benchmarks were successfully ported and evaluated. Several additional PARSEC benchmarks remain unported due to unresolved CHERI compatibility issues and are therefore left for future work.

The performance results in Figure~\ref{fig:parsec-overhead} show that PoisonCap introduces no runtime overhead compared with Cornucopia Reloaded for either single-threaded or two-thread executions. Although PoisonCap exhibits small performance improvements on several benchmarks, it is marginally slower on certain benchmarks such as \textit{freqmine}, where the slowdown is attributable to the additional CHERI tag cache pressure and the increased revoker workload caused by the increased presence of poison capabilities in memory. Furthermore, the two-thread results indicate that PoisonCap does not adversely affect multicore scaling relative to Cornucopia Reloaded. Nevertheless, the results should be interpreted in the context of the evaluation platform's limitations. As the FPGA prototype supports only two hardware threads, the study cannot fully characterize scalability on systems with larger core counts. Consequently, evaluating scalability on higher-core-count systems remains future work.   

\section{Related Work}

\textbf{Use-after-reallocation mitigation.}
PoisonCap is built on top of Cornucopia Reloaded, the existing CHERI temporal safety solution.
Reloaded delays memory reallocation until a memory sweep has revoked dangling capabilities~\cite{Cornucopia_reloaded}.
Reloaded requires memory zeroing to ensure safe reuse and lacks strict use-after-free mitigation support.
PoisonCap directly improves upon Reloaded with zeroing by enforcing hierarchical use-after-free detection and initialization safety, as well as communicating memory disuse to the cache subsystem.

Software-based approaches have also been proposed for temporal memory safety.
DangNull prevents dangling pointer dereferences by automatically nullifying pointers during object deallocation~\cite{dangnull}, while CETS maintains capability metadata to validate pointer dereferences at runtime~\cite{cets}. Modern managed-memory systems eliminate dangling pointers through automatic object lifetime management. An example is Oilpan~\cite{oilpan}, the garbage collector used by Chromium's Blink engine. 

\textbf{Memory tagging. }
ARM MTE mitigates use-after-free violations~\cite{ARM_MTE} but only provides probabilistic detection. Its limited tag size (e.g., 16 unique tags) leads to inevitable tag reuse across allocations without quarantine and does not guarantee temporal safety, as pointer arithmetic can be used to move a pointer to another allocation of the same type or, indeed, to change the pointer type. Apple's enhanced memory tagging solution (EMTE) reduces overhead compared to ARM's original MTE~\cite{nohARMMTEPerformance2026}. 
However, these memory tagging solutions are ill-suited for mitigating use-after-free violations in the nested allocation layers.
MTE supports a single tag for each memory word, which does not allow hardware or software to selectively allow access at various levels of an allocator hierarchy. 

\textbf{Hardware table-based solutions. }
CHERIoT is a CHERI-RISC-V specification and implementation for embedded applications~\cite{CHERIoT}.
Like Cornucopia, CHERIoT performs batch revocation by sweeping memory to revoke dangling capabilities; unlike Cornucopia, CHERIoT introduces a hardware load filter table that tracks freed memory locations.
CHERIoT checks every capability-wide memory access against the filter table (à la the shadow bitmap) to prevent loading dangling capabilities, thereby mitigating use-after-free violations.
This filter table is eminently practical when all memory is on-chip, as is typical for embedded systems.
However, this additional table lookup poses a far greater challenge for application-class CHERI systems with complex memory hierarchies backed by large DRAM memories.

Picasso proposes a table-based temporal safety solution that is more practical for application-class CHERI systems~\cite{gulmezPICASSOScalingCHERI2026}.
Picasso introduces an object validity table with one bit per allocation that is looked up on every heap memory access and can be efficiently cached in the load/store unit.
Picasso also enables strict use-after-free mitigation and reduces the frequency of revocation passes.
However, Picasso requires a large system-wide object table shared across CPU cores, which introduces significant synchronization costs.

In contrast, PoisonCap eliminates table lookup and instead employs a hierarchical, decentralized approach to use-after-free mitigation, avoiding adding the performance and memory overheads.

\textbf{Memory black-listing. }
Califorms supports byte-granularity black-listing by embedding metadata inline with user data to avoid memory overhead~\cite{sasakiPracticalByteGranularMemory2019}.
Compiler support is required to explicitly black-list bytes in padded regions, which triggers exceptions on access.
To improve compatibility with common compiler optimizations, Califorms needs to selectively suppress exceptions to preserve usability and compatibility. 
Califorms also treats bulk copy operations specially to allow the copying of black-listed bytes.

In contrast, PoisonCap carefully considers the delegation of memory poisoning privileges and naturally fits into existing software hierarchies without artificial allowances.

\textbf{Capabilities for initialization safety. }
``Mon CHÉRI'' added \emph{conditional capabilities} to CHERI to enforce a compiler-driven write-before-read policy for heap and stack initialization safety~\cite{gulmezMonCHERIMitigating2025}.
Mon CHÉRI requires substantial program transformation; maintaining consistency between conditional capabilities in register files necessitates a store linearization compiler pass.
This requirement means that applicability to large, real-world programs remains unexplored, and its performance overhead and compatibility are unclear.

In contrast, PoisonCap does not require such transformations, tracking state in the data itself rather than in capability references.
The PoisonCap approach should have no barriers when scaling to large software corpora, as macOS supports heap initialization safety using a similar, fixed data pattern~\cite{macos}.
PoisonCap improves on data pattern initialization by using unforgeable poison capabilities with reliable, un-bypassable write-before-read protection.

\section{Future Work}

\textbf{Kernel temporal safety. }
Beyond user-space, operating system kernels have an even richer memory allocation hierarchy; for example, the UMA, vmem and vm\_map layers in FreeBSD.
Hierarchical revocation support is even more desirable in the kernel to avoid leaking dangling capabilities at multiple layers.
While we do not address kernel temporal violations in our threat model, we expect poison capabilities to scale to this use case in future work.

\textbf{Revoker optimization. }
The revoker may benefit from microarchitectural optimizations enabled by poison capabilities.
The poison probe may leverage an enhanced \emph{CGetPoison} instruction to determine whether a poison capability exists at the given location and whether its poison bounds fully contain a given capability.
This probe instruction could bypass the CLG faults altogether and simplify fault handling.
Note that such a \emph{CGetPoison} instruction would not invalidate the revocation invariant because it would not load the contents of memory into the register file; instead, it could just act as a boolean test.

\textbf{Poison-aware tag controller. }
The CHERI revoker can skip loading and examining poison capabilities for revocation if it has a poison-aware \emph{CLoadTags} instruction~\cite{cheriv9}. The \emph{CLoadTags} relies on the CHERI tag controller to examine whether a cache line contains a capability. The CHERI tag controller needs to be able to keep track of poison capabilities differently from the CHERI capability to give a non-poison \emph{CLoadTags} response. Extending the CHERI tag controller to be poison-aware can also minimize tag cache pressure and improve revocation performance.

\section{Conclusion}
PoisonCap is a novel software-hardware co-design that improves the CHERI temporal safety security model to support hierarchical strict use-after-free mitigation and initialization safety by storing poison capabilities in freed memory regions.
PoisonCap supports hierarchical use-after-free mitigation and
initialization safety without additional performance overhead relative to Cornucopia with memory zeroing.
We have produced full prototype implementations in QEMU and FPGA and have identified a number of possible microarchitectural and software performance optimizations for future work.
We have used these prototypes to enable detailed full-system
evaluations, including the Juliet test suite, demonstrating PoisonCap's use-after-free detection and initialization-safety mechanisms.
SPEC 2006 and PARSEC benchmark results indicate very low overhead for the strong protection provided.
As with other CHERI mechanisms, PoisonCap is software-driven and incurs no baseline overhead for software that does not use these features, enabling incremental adoption where it is found to be most beneficial.

\section*{Acknowledgments}
This work was sponsored by UKRI3001: CHERI Research Centre, and the University of Cambridge Harding Distinguished Postgraduate 
Scholars Programme. 

Distribution Statement A: Approved for public release; distribution is unlimited. This work is also sponsored in part by the Defense Advanced Research Projects Agency (DARPA) under contract FA8750-24-C-B047 (“DEC”) as part of the CPM program. The views, opinions, and/or findings contained in this report are those of the authors and should not be interpreted as representing the official views or policies, either expressed or implied, of the Department of Defense or the U.S. Government.

We thank our colleagues on the CHERI team from the University of Cambridge for building the CHERI research platforms, and especially Jessica Clark for her assistance in getting our experiments up and running. We are also grateful to Merve Gulmez and Thomas Nyman at Ericsson Security Research for their help in reproducing the evaluations they had previously conducted. 

{
\bibliographystyle{ACM-Reference-Format}
\balance
\bibliography{MyLibrary}
}

\end{document}